\documentclass[runningheads]{llncs}
\usepackage{graphicx}
\usepackage{hyperref}
\usepackage{csquotes}
\usepackage{breakcites}
\usepackage{todonotes}
\usepackage{chngpage}
\usepackage{caption}
\usepackage{subcaption}

\graphicspath{{./figures/}}

\renewcommand{\paragraph}[1]{\textbf{#1.}}
\newcommand{\kst}{Institut der Kasseler Stottertherapie}
\begin{document}
\title{Towards Automated Assessment of Stuttering and Stuttering Therapy}
%
%
\author{Sebastian P. Bayerl\inst{1} \and
Florian H\"onig\inst{2}
Jo\"elle Reister\inst{2} \and
Korbinian Riedhammer\inst{1}
}
\authorrunning{S. P. Bayerl et al.}
%
\institute{Technische Hochschule Nürnberg Georg Simon Ohm, \textsc{Germany}\and
\kst\\
\email{\{sebastian.bayerl,korbinian\}@ieee.org}}

%
\maketitle              
\begin{abstract}

    Stuttering is a complex speech disorder that can be identified by repetitions, prolongations of sounds, syllables or words and blocks while speaking.
    Severity assessment is usually done by a speech therapist.
    While attempts at automated assessment were made, it is rarely used in therapy.
    Common methods for the assessment of stuttering severity include percent stuttered syllables (\%\,SS), the average of the three longest stuttering symptoms during a speech task or the recently introduced Speech Efficiency Score (SES).
    This paper introduces the Speech Control Index (SCI), a new method to evaluate the severity of stuttering.
    Unlike SES, it can also be used to assess therapy success for fluency shaping.
    We evaluate both SES and SCI on a new comprehensively labeled dataset containing stuttered German speech of clients prior to, during and after undergoing stuttering therapy.
    Phone alignments of an automatic speech recognition system are statistically evaluated in relation to their relative position to labeled stuttering events.
    The results indicate that phone length distributions differ in respect to their position in and around labeled stuttering events.

\end{abstract}

\section{Introduction}\label{sc:intro}

Stuttering is a speech disorder with a prevalence of 1\,\% of the population \cite{Carlson2012}.
It is a complex disorder of nerve coordination between both brain hemispheres.
It can be identified by repetitions, prolongations of sounds, syllables or words, and blocks while speaking.

In addition to these so-called core symptoms, a wide variety of linguistic, physical, behavioral and emotional accompanying symptoms can occur, some of them overlapping the core symptoms.
Stuttered disfluencies are usually accompanied by physical tension \cite{lickley2017disfluency}.
The frequency of occurrence and the duration of the symptoms vary considerably depending on individual severity and can seriously impair the communication of the person who stutters (PWS) \cite{stotern_j_und_e_2017}.
The individual appearance of the symptoms of each PWS also depends on the respective communication situation, the linguistic complexity of the utterance and the typical phased progress of the speech disorder \cite{handbook_stuttering_2008} \cite{packman2004theoretical}.
Since PWS know exactly what they want to say, the cause of the stuttered disfluency does not lie in planning or formulating speech, but in executing the plan of articulation \cite{lickley2017disfluency}.
The condition is treatable but not curable.

One possible technique to overcome stuttering is a technique called fluency shaping \cite{ingham2001evaluation,webster1972operant}.
Good results could be achieved by adapting it to stuttering therapy \cite{mallard1982precision}.
PWS learn a method to overcome blocks which is characterized by "easy" voice onset \cite{voiceonset_1985}.
A German adaption of this technique is the \emph{Kasseler Stottertherapie} which has also been proven to work well \cite{kst_2000,kst_euler2009}.
To assess the severity of stuttering and stuttering therapy success in some way, it is important to measure stuttering in a reliable way.
This is important both for therapeutic practice and research.
A stuttering diagnosis consists of the objective and subjective evaluation of the stuttering symptoms as well as the evaluation of the impairment of everyday life caused by the disorder.
It should provide a reliable picture of the individual severity of stuttering.

The objective evaluation of linguistic symptoms typically measures the frequency of stuttering events in percent of stuttered syllables ( \%\,SS ), whereby the number of stuttered syllables is related to overall spoken syllables.
However, this measure has only little agreement among different observers \cite{handbook_stuttering_2008} and does not take into account the type of stuttering symptom, e.g. one-time syllable repetition vs. several-second tense block, nor its duration, which significantly reduces the significance of  \%\,SS regarding the severity of stuttering \cite{handbook_stuttering_2008,starkweather1987fluency}.
Additionally to  \%\,SS , the duration of stuttering events can be determined in order to increase the reliability of the results.
However, commonly only a small part of the duration of stuttering events is taken into account, e.g.; in SSI-4 only the  average of the three longest stuttering symptoms is used \cite{riley2009ssi}.
These methods also do not record atypical stuttering disfluencies, which however can occur as accompanying linguistic symptoms and can significantly influence the impression of the severity of stuttering. 
Subjective stuttering severity rating scales are a widely used measure for assessing the severity of stuttering. 
These are commonly used both in speech therapy \cite{lidcombe_2003_stuttering} and in clinical research \cite{yairi1999early}.
For clinical purpose, severity rating scales are more reliable than  \%\,SS , to provide a statement about individual stuttering severity \cite{karimi_2014_absolute_and_relative_reliability}.

Methods for the automated assessment of stuttering and stuttering severity have been proposed in the past.
Nöth, Niemann, Haderlein, et al. use a standard speech recognition system and and evaluated vowel and fricative durations on a standardized reading task to discriminate between PWS and normal speakers \cite{noth_stuttering_2000}.
To classify prolongations and repetitions, Chee et al. extracted Mel Frequency Cepstral Coefficients (MFCC) and used them to train k-NN and LDA classifiers on a very small sample taken from the University College London Archive of Stuttered Speech \cite{mfcc_stuttering_chee_2009}\footnote{Available at https://www.uclass.psychol.ucl.ac.uk/uclassfsf.htm}.
Mundada et al. use the K-Means clustering algorithm to separate normal speakers from PWS.
They also use MFCC feature extraction and Dynamic Time Warping (DTW) for classification \cite{recognition_stuttering_mundada_2014}. 
{\'S}wietlicka et al. use artificial neural networks to discriminate between syllable repetitions, blocks before words that start with a plosive, and phone prolongations \cite{ann_stuttering_swietlicka_2013}.
Alharbi et al. recognize the need to develop customized ASR that can produce full verbatim transcripts including pseudo words and word parts without meaning.
Their approach is mainly focused on the detection of repetitions
\cite{lightly_supervised_stuttering_alharbi_2018}.
Ochi et al. investigated the automatic evaluation of soft articulatory contact, as it is taught in stuttering therapy. Detecting modified speech is necessary to account for it in automatic evaluation of PWS that went through speech therapy
\cite{ochi_soft_articulartory_2018}.


\paragraph{Our Contributions}
In this work, we introduce the Speech Control Index (SCI), a new method to evaluate the severity of stuttering which can also be used to assess therapy success for fluency shaping.
We evaluate both SES and SCI on a new comprehensively labeled dataset acquired at the \kst (KST) containing stuttered German speech of clients prior to, during and after undergoing stuttering therapy.
Based on phone alignments of an automatic speech recognition system, we perform a statistical evaluation of phone length distributions in relation to stuttering events.

\section{Data}\label{sc:data}

The data used in this paper was specifically created and labeled with stuttering and stuttering therapy in mind.
In the future, data gathered for this work will be used to create means to provide unobstrusive monitoring of stuttering.
Thus, the dataset was created to represent reality as good as possible.
No special recording equipment was used and the dataset was recorded with consumer hardware.
All recordings were created before, during and after therapy at the KST.
The therapy contains a number of different tasks such as reading, calling unacquainted people for inquiry purposes or talking to strangers in the street.

The labeling was done by two clinical linguists familiar with stuttering therapy at the KST.
The data is labeled in great detail differentiating twelve states of fluent or disfluent speech as well as prosodic pauses and blocks.
The focus is to comprehensively label stuttering behavior such as interrupted or repeated words or sentences in whole or parts.
The dataset also labels interjections, which can be a typical stuttering related behavior, even though it is also common in regular speakers.
Another unique feature of the dataset is the labeling of modified speech: speech as it is produced when applying the fluency shaping technique taught and trained at the KST.
Additionally to the labeling of stuttering behavior, a transcript is provided in which word abortions are marked and transcribed in a verbatim way.
During preprocessing, the recordings were resampled to 16~kHz where necessary and in case of stereo recordings only one channel is used.
The dataset contains 214 recordings by 37 speakers of which 28 were male and 9 were female.
The dataset amounts to about 207 minutes of labeled speech.

To the best of our knowledge, these features make it one of the largest and most comprehensively labeled datasets containing stuttered speech.
One of its most important features is the existence of stuttered and modified speech prior to, during and after therapy, enabling extensive research and the creation of practical applications that can be used in a therapeutic context.

\section{Method}\label{sc:method}
To assess the severity of stuttering or disfluency of speakers using fluency shaping, common evaluation methods such as \%SS or SSI-4 are insufficient, as these methods do not account for therapy artefacts and accompanying linguistic symptoms.
Since a purely subjective measure of stuttering severity has many drawbacks in clinical practice, we chose to calculate the SES based on classifying speech as either \emph{efficient}, \emph{inefficient} or \emph{silence}.

\subsection{Speech Efficiency Score}
The Speech Efficiency Score (SES) is a recent method for the evaluation of (dis)fluent speech that was proposed by Amir et al. \cite{speech_efficiency_Amir2018}.
This method puts the fraction of fluent speech in relation to the fraction of disfluent speech.
Thus, SES determines the communicative efficiency of a speaker by focusing on the time domain.

\begin{equation}\label{eq:ses}
    SES =  \mathrm{\frac{Efficient\ time}{Total\ time - Silence}} \cdot 100\%
\end{equation}

With this method, all kinds of disfluencies, both typical and atypical to stuttering, are taken into account, as well as the duration of the fluent and disfluent speech components, which makes it superior to previous methods.
Amir et al. concluded that, due to the high correlation they found between SES and subjective severity rating scales, SES also provides reliable information about the severity of stuttering.
Since SES considers prolongations, which are perceived as abnormal, to be inefficient, it must be assumed that the SES fails to take adapted speaking behaviors into account.
This in turn implies that for the calculation of SES, speech fractions that contain modified speech, such as fluency shaping, are counted as \emph{inefficient}.

Fluency shaping focuses on restructuring the way of speaking, aiming at modifying speech in a way that little or no stuttering symptoms occur.
The technique includes gentle voice onsets as well as syllable and word bindings, in which the vibration of the vocal cords is not supposed to stop.
It allows PWS to regain a high degree of control over their own speech and speak much more fluently.
However, applying this technique, especially at the beginning of the therapy, sounds quite unnatural due to the prolongations that are not present in a normal flow of speech \cite{packman_1994_prolonged}.
Since calculating SES includes speech fractions that have been modified by fluency shaping as \emph{inefficient}, the measure does not give a reliable picture of the severity of stuttering in PWS who apply this technique.

\subsection{Speech Control Index}

To address the shortcomings of SES in the context of speech therapy using fluency shaping, we propose a new method that can be used to assess the severity of stuttering but still is able to account for and measure therapeutic success.
The Speech Control Index (SCI) was developed at the KST and accounts for speech modifications which relate to fluency shaping.
By adding modified speech to the controlled speech, the SCI not only provides a measure for the individual severity of stuttering, but also whether or not PWS are able to control their speech by using the speaking technique.
The SCI quantifies the proportion of time between \emph{controlled} speech components, which means fluent and modified speech, and \emph{uncontrolled} speech components such as disfluencies and blocks.
Thus SCI, similar to SES, considers speaking over time.

To achieve this, speech fractions are grouped in one of three categories:
\begin{enumerate}
    \item \textbf{Controlled time} - all parts of speech produced that can be considered fluent or modified, which means a PWS uses a speaking technique to overcome stuttering. Additionally prosodic pauses are added to this category.
    \item \textbf{Disfluent time} - all parts of a sample that can be identified as stuttered disfluencies are being counted to disfluent speech, i.e.~repetitions of sounds, syllables, words, prolongations, blocks and silent blocks.
    In addition, speech fractions containing atypical stuttering disfluencies such as the repetition of phrases, interjections, revisions including incomplete words and phrases are being added to disfluent time.
    \item \textbf{Silence} - long pauses in which the PWS is not speaking and not trying to speak as well as interruptions by the dialogue partner, etc.

\end{enumerate}

Accordingly, "Total time" in Eq.~\ref{eq:sci} is the sum of the three aforementioned categories.

Based on the correlation between subjective severity rating scales and the SES, Amir et al. concluded that SES also provides reliable information about the severity of stuttering \cite{speech_efficiency_Amir2018}.
As calculation of SCI is similar to SES beside the attribution of modified speech fractions, the same is expected to hold for the SCI.
In cases where PWS do not use the speaking technique, which can be assumed for recordings done prior to therapy, both measures are equal.
The same is true for cases in which only little speaking technique is applied, which is confirmed by Fig.~\ref{fig:ses_sci_plot}.

\begin{equation}\label{eq:sci}
    SCI =  \mathrm{\frac{Controlled\ time} {Total\ time - Silence}} \cdot 100~\%
\end{equation}


\subsection{Phone Durations}

One of the core symptoms of stuttering is the prolongation of sounds.
This should be directly observable in the time alignment outputs produced by an automatic speech recognition (ASR) system.
Such information can be used to differentiate between a PWS and a normal speaking person.
A major difficulty is that phone lengths are unique speech properties characteristic of every speaker and may vary depending on various factors.
To generalize such an assessment, a sample of multiple speakers is necessary.
It can be assumed that especially close to and during a stuttering event, phone durations should on average be longer than during fluent speech portions.
To verify these assumptions, phone alignments were produced and categorized with respect to their relative position to stuttering events:
Phones inside labeled disfluencies, phones within 0.25 seconds before a disfluency, 0.25 seconds before and after a disfluency, 0.25 seconds before, after and inside a disfluency.
To have a set that is free from modifications, which also prolong phone lengths, a set of phones was chosen which where within speech fractions labeled as fluent.
The sets were then refined by the phone classes vowels, fricatives, sonorants and plosives.
Altogether, 44 sets of phone duration distributions were created, but the individual sets became to small to make generalizable conclusions.

To obtain the alignments for calculating phone lengths, an ASR system trained based on the system described in \cite{milde_2018} was used.
For training, the German part of the Spoken Wikipedia Corpora, the German subset of the m-ailabs read speech corpus as well as the Tuda-De corpus were used \footnote{Kaldi recipe available at https://github.com/uhh-lt/kaldi-tuda-de}.
Only minor modifications to the training recipe were made to reduce the number of training targets in acoustic model training from 732 to 260.
The model was trained using the Kaldi toolkit \cite{kaldi2011}, using  speaker adaptive training on top of LDA and MLLT features \cite{fast_sat_povey_2008}.
Prior to computing the forced alignments, the lexicon transducer of the ASR system was modified to be able to align incomplete words.
The transcripts created for the files were checked against the lexicon and pronunciations for missing and incomplete words were generated by using a grapheme-to-phoneme (g2p) model trained on the original lexicon\footnote{G2P tool available online at https://www-i6.informatik.rwth-aachen.de/web/Software/g2p.html}.

\section{Experiments}\label{sc:experiments}

\begin{figure}
    \begin{adjustwidth}{-1.4in}{-1in}%

    \centering
        \begin{subfigure}[t]{.8\textwidth}
            \includegraphics[width=.8\linewidth]{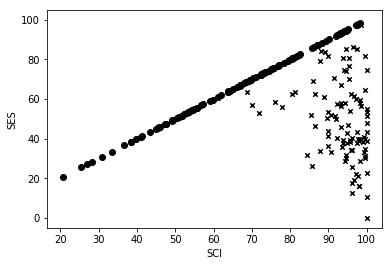}
            \caption{Plot of SES over SCI values computed from labels for every file in the dataset. Crosses representing samples that contain modified speech, dots representing samples without.}
            \label{fig:ses_sci_plot}
        \end{subfigure}%
        \begin{subfigure}[t]{.8\textwidth}
            \centering
            \includegraphics[width=.8\linewidth]{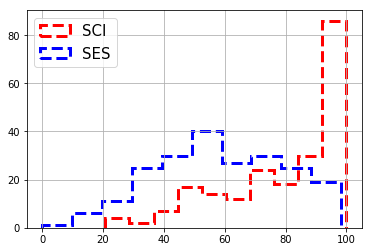}
            \caption{Value distribution of SCI and SES scores in the dataset ($N = 214$).}
            \label{fig:sci_ses_value_dist}
        \end{subfigure}

    \end{adjustwidth}%

\end{figure}

SCI and SES were computed for each of the 214 files in the dataset.
Pearson's correlation between the SCI and SES over all 214 files is at 0.142 and only shows a very weak linear relationship between the two indices.
This is confirmed by the distribution plots in Fig.~\ref{fig:sci_ses_value_dist}, and the irregular plot for SES values over higher SCI values in Figure \ref{fig:ses_sci_plot}.
Comparing the SCI and SES directly, the absolute difference is less than 0.1 percentage points for 114 recordings, which is indicated by the plot in Figure \ref{fig:ses_sci_plot}.
This is exactly the part of the data that has no labeled modifications in it, which is supported by a correlation of 1 between SCI and SES for this part of the data.
This shows that SCI and SES are identical for samples without speech modifications.

\begin{figure}[t]
    \centering
    \includegraphics[width=\linewidth, height=7cm,keepaspectratio]{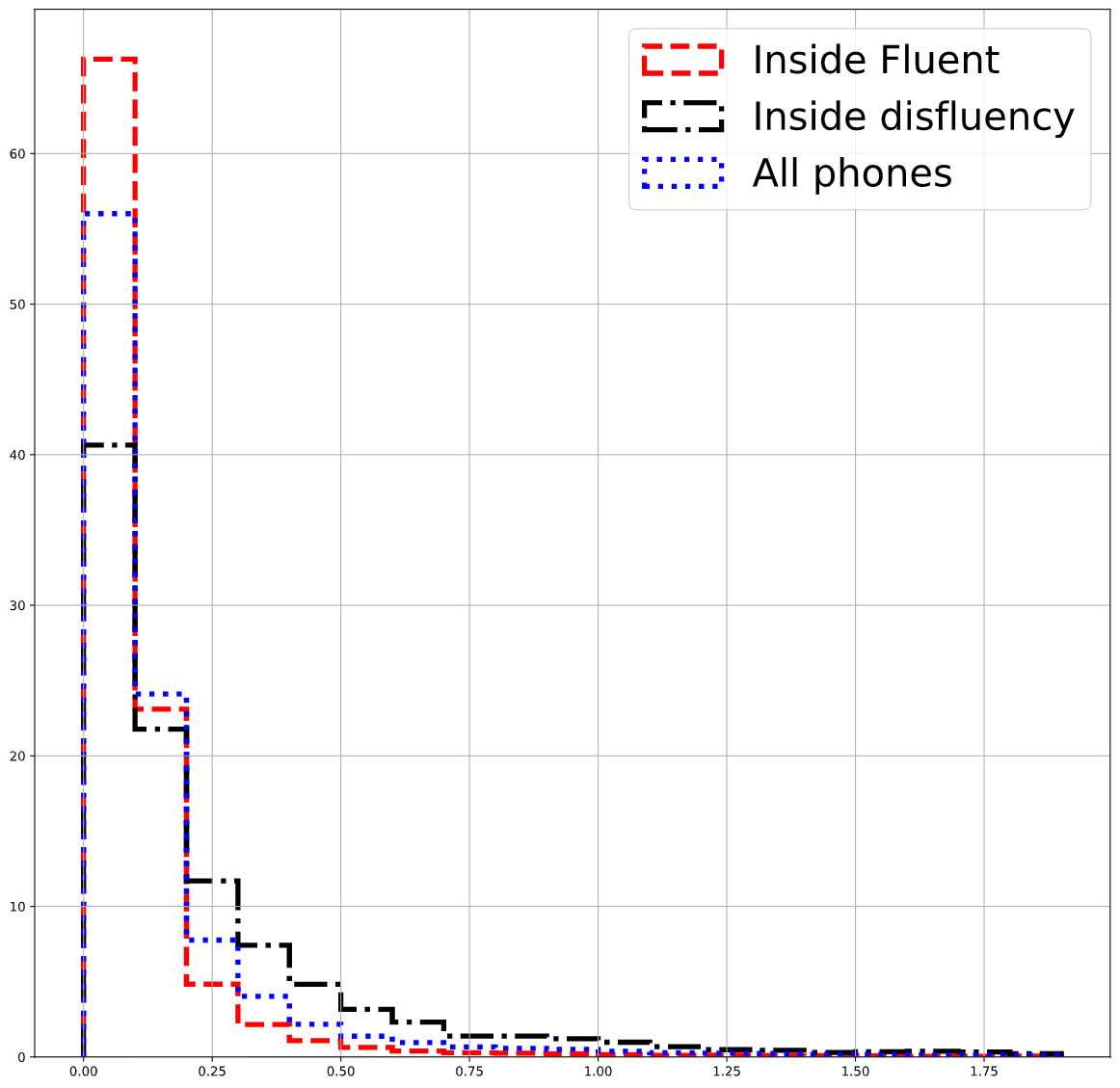}
    \caption{Fluent, All, and inside disfluency phone length distributions plotted as relative portions of phone durations. Area under step function represents percentage of values inside a 0.1 second wide phone length interval.}
    \label{fig:relative_dist}
\end{figure}%

    \begin{table*}[tb]
        \begin{adjustwidth}{-1in}{-1in}%

        \caption{\label{tbl:dist}
        Phone duration (in seconds) distributions descriptive statistics.}
        
        \centering
        \begin{tabular}{|l|r|c|c|c|c|}
            \hline
            \textbf{Dataset}         & \textbf{N} & \textbf{Mean phone dur.} & \textbf{Phone dur. at 90\textsuperscript{th} P} & \textbf{Phone dur. at 95\textsuperscript{th} P} & \textbf{Percent Outlier} \\ \hline
            Inside disfluency             & 9818       & 0.230                    & 0.570                         & 0.850                         & 3.14                    \\ \hline
            Before disfluency              & 7898       & 0.199                    & 0.460                         & 0.670                         & 2.12                    \\ \hline
            Before to after disfluency     & 23227      & 0.192                    & 0.480                         & 0.730                         & 2.58                    \\ \hline
            All phones               & 73410      & 0.150                    & 0.330                         & 0.520                         & 1.99                    \\ \hline
            Fluent                   & 41195      & 0.109                    & 0.200                         & 0.310                         & 0.84                    \\ \hline
        \end{tabular}
    \end{adjustwidth}

    \end{table*}

Tab.~\ref{tbl:dist} shows the descriptive statistics about the created phone subsets.
In this context, outliers were defined as phones of which the duration is at least three times the standard deviation $\sigma$ greater than the mean phone duration in the overall set.
The set containing all phones has about 2\% outliers, which is higher than the expected value for this definition of outliers.
The difference to the set containing only fluent speech as well as the set inside labeled disfluencies is most striking.
Fluent speech only contains 0.84\% outliers and the average phone duration relative to the set containing all phones is 27\% shorter.
Phones inside a disfluency compared to the set containing all phones are on average 53\% longer.
Relative difference between average duration of phones inside fluently labeled speech compared to speech inside disfluencies is 111\%.
These numbers show a clear relationship between phone duration and stuttering related disfluencies.

It can be concluded that especially phone durations starting from the 90\textsuperscript{th} percentile within a sample can be very useful in differentiating stuttered speech from normal speech.
The plot in Fig.~\ref{fig:relative_dist} supports this observation.
It contains histogram plots of the relative portion of phone durations in 0.1 second wide intervals.
This shows that apart from the phones with a duration below 0.2 seconds, the relative  number of phones inside these 0.1 second wide intervals is greatest for phones inside disfluencies.
Looking at the relative fraction of phones above or below a duration of 0.2 seconds, might be enough to differentiate between fluent and disfluent speech.

\section{Conclusion}\label{sc:conclusion}

The SCI provides an accurate measure with similar properties as the SES for speakers who speak mostly fluent or do not use a special speech technique.
The advantage of SCI is its ability to account for modified speech of PWS who underwent therapy and regained a level of fluency and control that is more effective than stuttering, even though speech may not be classified as natural or normal.
An extensive comparison between the objective measures SSI-4 ( \%\,SS and mean duration of the three longest symptoms), SES and SCI, as well as a comparison of these procedures with subjective stuttering severity rating scales will be a part of future work.

The data showed that there is a clear relation between the duration of phones and their relative position to stuttering events.
As indicated here, a normal speech recognition system can be easily modified to distinguish fluent and disfluent speech in utterances based on heuristic measures as long as it can produce alignments.
For this the recognition system needs to be able to recognize incomplete words and syllable repetitions.
This insight will be used to build automatic stuttering recognition systems that can differentiate different levels of fluency.
The comprehensively labeled dataset enables future exploration of different kinds of disfluencies and the use of statistical learning methods such as support vector machines or neural networks.
By classifying the amount of fluent, disfluent and modified speech in a speech sample, the automated and continuous calculation of the SCI can provide a reliable measure for stuttering severity and therapy success.
This will provide valuable feedback to the client as well as the therapist.

\section{Acknowledgements}

The authors thank the \kst \ for their support and excellent collaboration.
This work is supported by a research grant of the Bayerisches Staatsministerium für Bildung und Kultus, Wissenschaft und Kunst as well as the BAYWiss (Bayerisches Wissenschaftsforum).

%
%
%

%
%
%
%
\bibliographystyle{splncs04}
\bibliography{0_paper.bib}

\begin{thebibliography}{10}
\providecommand{\url}[1]{\texttt{#1}}
\providecommand{\urlprefix}{URL }
\providecommand{\doi}[1]{https://doi.org/#1}

\bibitem{lightly_supervised_stuttering_alharbi_2018}
Alharbi, S., Hasan, M., Simons, A.J., Brumfitt, S., Green, P.: A lightly
  supervised approach to detect stuttering in children's speech. In:
  Proceedings of Interspeech 2018. pp. 3433--3437. ISCA (2018)

\bibitem{speech_efficiency_Amir2018}
Amir, O., Shapira, Y., Mick, L., Yaruss, J.S.: {The Speech Efficiency Score
  ({SES}): A time-domain measure of speech fluency}. Journal of Fluency
  Disorders  (aug 2018). \doi{10.1016/j.jfludis.2018.08.001}

\bibitem{stotern_j_und_e_2017}
Anders, K., Rudorf, E.: Kompendium der Akademischen Sprachtherapie und
  Logopädie: Bd.3, chap. Stottern bei Jugendlichen und Erwachsenen., pp. {225
  -- 241}. Kohlhammer W. (2017)

\bibitem{Carlson2012}
Carlson, N.R.: Physiology of Behavior (11th Edition). Pearson (2012)

\bibitem{mfcc_stuttering_chee_2009}
Chee, L.S., Ai, O.C., Hariharan, M., Yaacob, S.: {MFCC Based Recognition of
  Repetitions and Prolongations in Stuttered Speech using k-NN and LDA}. In:
  2009 IEEE Student Conference on Research and Development (SCOReD). pp.
  146--149. IEEE (2009)

\bibitem{handbook_stuttering_2008}
Ellis, J.B., Ramig, P.R.: Journal of Fluency Disorders  \textbf{34}(4),  295 --
  299 (2008). \doi{https://doi.org/10.1016/j.jfludis.2009.10.004}

\bibitem{kst_euler2009}
Euler, H.A., v.~Gudenberg, A.W., Jung, K., Neumann, K.: {Computergestützte
  Therapie bei Redeflussstörungen: Die langfristige Wirksamkeit der Kasseler
  Stottertherapie ({KST})}. Sprache $\cdotp$ Stimme $\cdotp$ Gehör
  \textbf{33}(04),  193--202 (dec 2009). \doi{10.1055/s-0029-1242747}

\bibitem{kst_2000}
Euler, H., Wolff~v. Gudenberg, A.: {Die Kasseler Stottertherapie (KST).
  Ergebnisse einer computer-gestützten Biofeedbacktherapie für Erwachsene*1}.
  Sprache-stimme-gehor  \textbf{24},  71--79 (06 2000).
  \doi{10.1055/s-2000-11084}

\bibitem{ingham2001evaluation}
Ingham, R.J., Kilgo, M., Ingham, J.C., Moglia, R., Belknap, H., Sanchez, T.:
  Evaluation of a stuttering treatment based on reduction of short phonation
  intervals. Journal of Speech, Language, and Hearing Research  \textbf{44}(6),
   1229--1244 (2001)

\bibitem{voiceonset_1985}
J.~Borden, G., Baer, T., Kay~Kenney, M.: Onset of voicing in stuttered and
  fluent utterances. Journal of speech and hearing research  \textbf{28},
  363--72 (10 1985). \doi{10.1044/jshr.2803.363}

\bibitem{karimi_2014_absolute_and_relative_reliability}
Karimi, H., O’Brian, S., Onslow, M., Jones, M.: Absolute and relative
  reliability of percentage of syllables stuttered and severity rating scales.
  Journal of Speech, Language, and Hearing Research  \textbf{57}(4),
  1284--1295 (2014)

\bibitem{lickley2017disfluency}
Lickley, R.: Disfluency in typical and stuttered speech. Fattori sociali e
  biologici nella variazione fonetica-Social and biological factors in speech
  variation  (2017)

\bibitem{mallard1982precision}
Mallard, A., Kelley, J.: The precision fluency shaping program: Replication and
  evaluation. Journal of Fluency Disorders  \textbf{7}(2),  287--294 (1982)

\bibitem{milde_2018}
Milde, B., K{\"o}hn, A.: Open source automatic speech recognition for {German}.
  In: Proceedings of ITG 2018 (2018)

\bibitem{recognition_stuttering_mundada_2014}
Mundada, M., Gawali, B., Kayte, S.: Recognition and classification of speech
  and its related fluency disorders. International Journal of Computer Science
  and Information Technologies (IJCSIT)  \textbf{5}(5),  6764--6767 (2014)

\bibitem{noth_stuttering_2000}
N{\"o}th, E., Niemann, H., Haderlein, T., Decher, M., Eysholdt, U., Rosanowski,
  F., Wittenberg, T.: Automatic stuttering recognition using hidden markov
  models. In: Sixth International Conference on Spoken Language Processing
  (2000)

\bibitem{ochi_soft_articulartory_2018}
Ochi, K., Mori, K., Sakai, N.: Automatic evaluation of soft articulatory
  contact for stuttering treatment. Proc. Interspeech 2018 pp. 1546--1550
  (2018)

\bibitem{lidcombe_2003_stuttering}
Onslow, M., Packman, A., Harrison, E., et~al.: The Lidcombe Program of Early
  Stuttering Intervention: A Clinician's Guide. Pro-ed Austin, TX (2003)

\bibitem{packman2004theoretical}
Packman, A., Attanasio, J.S.: Theoretical issues in stuttering  (2004)

\bibitem{packman_1994_prolonged}
Packman, A., Onslow, M., Doorn, J.v.: Prolonged speech and modification of
  stuttering: Perceptual, acoustic, and electroglottographic data. Journal of
  Speech, Language, and Hearing Research  \textbf{37}(4),  724--737 (1994)

\bibitem{kaldi2011}
Povey, D., Ghoshal, A., Boulianne, G., Goel, N., Hannemann, M., Qian, Y.,
  Schwarz, P., Stemmer, G.: The {Kaldi} speech recognition toolkit. In: In IEEE
  2011 workshop (2011)

\bibitem{fast_sat_povey_2008}
Povey, D., Kuo, H.K.J., Soltau, H.: Fast speaker adaptive training for speech
  recognition. In: Ninth Annual Conference of the International Speech
  Communication Association (2008)

\bibitem{riley2009ssi}
Riley, G.: {SSI-4 Stuttering Severity Instrument, Fourth Edition} (2009)

\bibitem{starkweather1987fluency}
Starkweather, C.W.: Fluency and stuttering. Prentice-Hall, Inc (1987)

\bibitem{ann_stuttering_swietlicka_2013}
{\'S}wietlicka, I., Kuniszyk-J{\'o}{\'z}kowiak, W., Smo{\l}ka, E.:
  {Hierarchical ANN system for Stuttering Identification}. Computer Speech \&
  Language  \textbf{27}(1),  228--242 (2013)

\bibitem{webster1972operant}
Webster, R.L.: An operant response shaping program for the establishment of
  fluency in stutterers. final report.  (1972)

\bibitem{yairi1999early}
Yairi, E., Ambrose, N.G.: {Early Childhood Stuttering I: Persistency and
  Recovery Rates}. Journal of Speech, Language, and Hearing Research
  \textbf{42}(5),  1097--1112 (1999)

\end{thebibliography}
\end{document}